\documentclass[%
twocolumn,
superscriptaddress,
 amsmath,amssymb,
 prl,
]{revtex4-2}

\usepackage{graphicx}% Include figure files
\usepackage{dcolumn}% Align table columns on decimal point
\usepackage{xcolor}

\usepackage{bm}% bold math
\begin{document}

%\preprint{APS/123-QED}

\title{Biphoton entanglement of topologically-distinct modes}% Force line breaks with \\

\author{Cooper Doyle}
\affiliation{%
 Institute of Photonics and Optical Science (IPOS), School of Physics, The University of Sydney, Sydney, NSW 2006, Australia.
}%
\author{Wei-Wei Zhang}%
\affiliation{Max Planck Institute for the Science of Light, Staudtstraße 2, 91058 Erlangen, Germany}
\affiliation{Institute for Theoretical Physics II,  Friedrich-Alexander-Universit{\"a}t Erlangen-N{\"u}rnberg,  Staudtstraße 7, 91058 Erlangen, Germany}
\affiliation{Gusu Laboratory of Materials, 215123 Suzhou, China}

\author{Michelle Wang}
\affiliation{%
 Institute of Photonics and Optical Science (IPOS), School of Physics, The University of Sydney, Sydney, NSW 2006, Australia.
}%

\author{Bryn A. Bell}%
\affiliation{%
Department of Physics, Imperial College London, Prince Consort Rd., London SW7 2AZ, UK
}%
\author{Stephen D. Bartlett}
\affiliation{Centre for Engineered Quantum Systems, School of Physics, The University of Sydney, Sydney, NSW 2006, Australia}

\author{Andrea Blanco-Redondo}
\email{andrea.blanco-redondo@nokia-bell-labs.com}
\affiliation{%
 Nokia Bell Labs, 600 Mountain Ave., New Providence, NJ 07974, USA
}%

\date{\today}% It is always \today, today,

\begin{abstract}
The robust generation and manipulation of entangled multiphoton states on-chip has an essential role in quantum computation and communication. Lattice topology has emerged as a means of protecting photonic states from disorder but entanglement across different topologies remained unexplored. We report biphoton entanglement between topologically distinct spatial modes in a bipartite array of silicon waveguides. The results highlight topology as an additional degree of freedom for entanglement and open avenues for investigating information teleportation between trivial and topological modes.
\end{abstract}

\maketitle

%\tableofcontents

%\section{Introduction}

Since the detection of topological phases in condensed matter systems, first identified with the discovery of the quantum Hall effect \cite{Klitzing1980, Thouless1982}, topological photonics~\cite{Ozawa2019} has appeared as a rapidly growing field of study. Most topological photonics experiments leverage bulk-edge correspondence \cite{Ryu2002}, which explains the emergence of localized modes at the interface between two materials with different topological invariants. These so-called topological modes are robust against symmetry-preserving disorder and have energies that lie within the energy gap of the surrounding bulk materials \cite{Ozawa2019}. 

In the past decade, experiments with classical light have demonstrated the topological protection of edge modes at  microwave \cite{Wang2009} and optical frequencies \cite{Kraus2012, Rechtsman2014, Hafezi2013, Ozawa2019}. More recently this research has been extended into the quantum realm, where quantum properties of light, such as entanglement, come into play and are subject to topological protection \cite{Blanco-Redondo2020}. Using nanophotonic lattices, recent works have demonstrated edge propagation of single photons \cite{Barik2018, Wang2019b}, spectrally \cite{Mittal2018} and spatially robust generation and propagation of photon pairs  \cite{Blanco-Redondo2018}, topological protection of spatial  \cite{Wang2019} and polarization mode entanglement \cite{wang2019topologically}, and robust quantum interference \cite{Tombasco2018}.

Several studies have evidenced the role that nonlinearities can play in modifying the topological properties of a system, where nonlinear interactions can enable two-photon edge states in systems that are otherwise topologically trivial in the linear regime \cite{Gorlach2017, Stepanenko2020, Olekhno2020}. Other recent theoretical works have shown that topological protection of multiphoton entangled states can deteriorate more quickly than that of their classical or single-photon counterparts \cite{Bergamasco2019, Tschernig2020}.

In all the above cases, the photonic quantum states populate modes of the same topology. Here, we report biphoton entanglement of modes of different topology, introducing topology as an additional degree of freedom for entanglement. Entanglement in multiple-DOFs enables many advantages in quantum technologies such as more efficient Bell measurements, superdense coding, and the generation of multi-qubit entangled states \cite{MultipleDOF}. Our platform is a bipartite array of silicon-on-insulator waveguides \cite{Blanco-Redondo2016} in which nonlinearly generated photon pairs are created in a superposition of three co-localized  modes of two distinct topologies. The classical pump giving rise to this biphoton entanglement, however, populates exclusively modes of the same topology, highlighting a striking difference in this system between the (linear) classical and (nonlinear) quantum regimes. 

\begin{figure}
\centering
\includegraphics[width=\columnwidth]{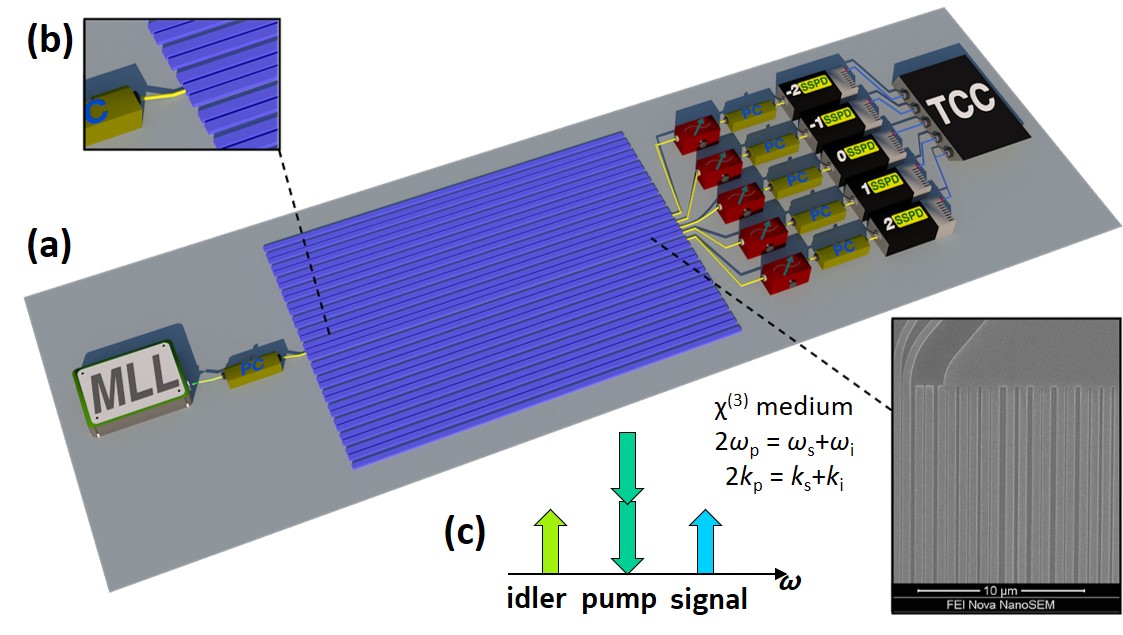}
\caption{CMOS-compatible bipartite array of silicon waveguides and experimental setup. (a) Pulses from a mode-locked laser are coupled into the center of a short-short defect of the array, whose five center waveguides are fanned out and coupled to individual tunable filters and polarization controllers before being detected with single-photon detectors and a time-correlation circuit. Inset: Scanning electron microscope (SEM) image of a portion of the array; (b) Detail of the short-short defect; (c) Schematic of degenerate spontaneous four-wave mixing.} 
\label{ExpSketch}
\end{figure}

The setup of our experiment is illustrated in Fig.~\ref{ExpSketch}~(a). The focus is an array of coupled silicon waveguides containing 203 silicon nanowires with length $L = 381 \ \mu m$, height $h = 220 \ nm$ and width $w = 450 \ nm$ on a silica substrate. We index these waveguides as [-101,101]  with the center waveguide indexed as 0. The air gaps separating the waveguides alternate between short gap $g_s = 173 \ nm$ and long gap $g_l = 307 \ nm$, in analogy to the Su-Schrieffer-Heeger (SSH) model \cite{Su1979}, and feature a short-short defect (two consecutive short gaps) around waveguide 0, shown in Fig.~\ref{ExpSketch}~(b). The short-short defect results in an interface between two mirror images of the SSH lattice, yielding a topological mode localized at this interface, as well as two topologically trivial modes localized by the region of higher refractive index \cite{Blanco-Redondo2016}. Picosecond laser pulses at $1550 \ nm$, with a $\sim150$ GHz bandwidth, are pumped into the center waveguide using a mode-locked laser (MLL). The high pulse peak power combined with strong spatial confinement in the silicon waveguides leads to the stochastic generation of correlated signal and idler photons via spontaneous four-wave mixing (SFWM), as schematically depicted in Fig.~\ref{ExpSketch}~(c). At the output, signal and idler photons are filtered using arrayed waveguide gratings and tunable filters at $\sim1545$ nm and $\sim1555$ nm respectively, which satisfies the conservation of energy condition imposed by SFWM. The signal and idler photons from the 5 central waveguides are then detected using cryogenically-cooled superconducting nanowire single-photon detectors (SSPDs), and their arrival times are matched using a time-correlation circuit (TCC), allowing us to map the spatial profile of the biphoton states in the lattice.

\paragraph{Classical and quantum dynamics. --}
The silicon waveguide lattice detailed above can be described by the Hamiltonian
\begin{gather}
H = H_\text{p} + H_\text{s} + H_\text{i} + H_\text{NL} \\
H_\text{p,s,i} = \sum_n t_n^\text{p,s,i} a_n^\text{p,s,i} a_{n+1}^{\text{p,s,i}\dagger} + t_{n}^\text{p,s,i} a_{n+1}^\text{p,s,i} a_{n}^{\text{p,s,i}\dagger} \\
% H_\text{s} = \sum_n t_n^\text{s} a_n^\text{s} a_{n+1}^{\text{s}\dagger} + t_n^\text{s} a_{n+1}^\text{s} a_n^{\text{s}\dagger} \\
% H_\text{i} = \sum_n t_n^\text{i} a_n^\text{i} a_{n+1}^{\text{i}\dagger} + t_n^\text{i} a_{n+1}^\text{i} a_n^{\text{i}\dagger} \\
H_\text{NL} = \gamma \sum_n a_{n}^{p2}a_n^{\text{s}\dagger}a_n^{\text{i}\dagger} + a_{n}^{p\dagger 2}a_n^\text{s}a_n^\text{i}
\label{eq:H-NL}
\end{gather}

\noindent
where $H_\text{p,s,i}$ represent the pump, signal, and idler linear Hamiltonians respectively, with $t_n^\text{p,s,i}$ being the coupling strength between waveguide $n$ and $n+1$ at the pump, signal, and idler frequencies. Operators $a_n^{\text{p,s,i}\dagger}$ and $a_n^{\text{p,s,i}}$ are the respective creation and annihilation operators at pump, signal, and idler frequencies. The term $H_\text{NL}$ represents the nonlinear Hamiltonian describing SFWM, with $\gamma$ being the nonlinear gain. For a sufficiently bright pump in the undepleted approximation, where the pump is unaffected by the pair creation, we can write $a \in \mathbb{C}^N$ as a vector of classical complex pump amplitudes.

\begin{figure}
\centering
\includegraphics[width=\columnwidth]{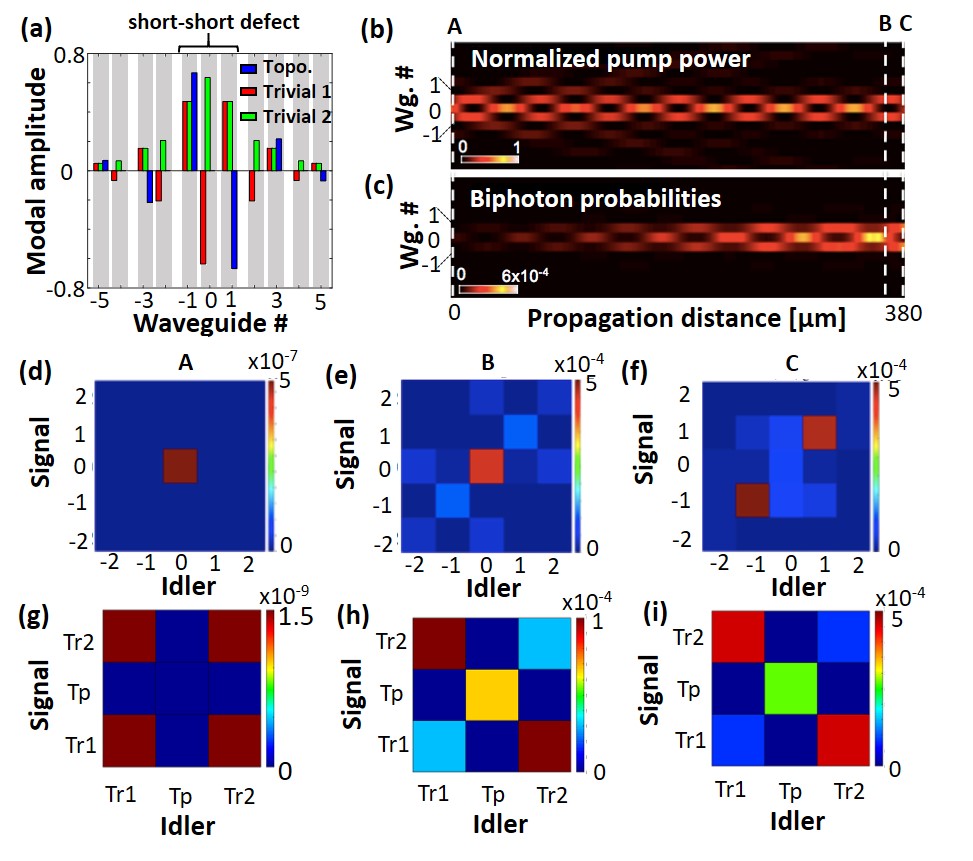}
\caption{Simulations of classical and quantum dynamics. (a) Amplitude distributions of the three
eigenmodes co-localized at the short-short defect; (b) Normalized pump intensity, showing beating
between two trivial modes; (c) Biphoton probabilities, showing beating between the
topological and the two trivial modes; (d-f) Spatial biphoton probabilities and (g-i) biphoton population of localized eigenmodes at propagation points A-C, respectively.}
\label{Dynamics}
\end{figure}

The lattice supports 200 eigenmodes extended across the lattice, as well as three localized modes which coexist at the short-short defect in the center of the nanophotonic lattice, as illustrated by their exponentially decaying modal amplitudes in Fig.~\ref{Dynamics}~(a): a topological boundary state (blue bars) – arising at the boundary because of the bulk-boundary correspondence – and two trivial modes (red and green bars) – localized by the region of higher refractive index at the short-short defect. Notice that the center waveguide of the defect has high support for the two trivially localized modes but no support for the topological boundary state. Therefore the pump injected at the center of the defect, depicted in Fig.~\ref{ExpSketch}~(a) and~(b), does not couple to the topological mode, only the trivial modes.

The linear part of the Hamiltonian can be rewritten in terms of the eigenmodes as
\begin{equation}
    H_\text{p,s,i}=\sum_j \Omega_j^{p,s,i} b_j^{p,s,i\dagger}b_j^{p,s,i}
\end{equation}
where $b_j^{p,s,i}$ is the annihilation operator for a pump, signal, or idler photon in the $j$th eigenmode, with eigenvalue $\Omega_j^{p,s,i}$. Similarly, the nonlinear Hamiltonian can be rewritten as:
\begin{equation}
    H_\text{NL}=\sum_{j,k,l,m} \Gamma_{jklm}b_j^pb_k^pb_l^{s\dagger}b_m^{i\dagger}+\Gamma^*_{jklm}b_j^{p\dagger}b_k^{p\dagger}b_l^{s}b_m^{i}
\end{equation}
with
\begin{equation}
    \Gamma_{jklm}=\gamma \sum_n f_{j,n}^pf_{k,n}^pf_{l,n}^{s*}f_{m,n}^{i*},
\end{equation}
where $f_{j,n}^{p,s,i}$ are the amplitudes for the $j$th eigenmode in the $n$th waveguide. This shows that SFWM can occur between any combination of eigenmodes, but depends on the overlap between the four mode-profiles, which will be largest for the localized modes. This overlap will however be zero for SFWM involving an odd number of anti-symmetric modes - so for example two pump photons from a trivial symmetric mode can be removed to create two photons in the anti-symmetric topological mode, but not to create one trivial and one topological photon.

Using an undepleted pump approximation, the classical pump amplitudes $a_n$ follow the linear dynamics equation
\begin{equation}
    \partial_z a_n=t_n^\text{p} a_{n+1}+t_{n-1}^\text{p} a_{n-1},
\end{equation}

\noindent
where $z$ is displacement in the direction of propagation. The coupling constants $t_\text{short}^\text{p}= 45044 \ m^{-1}$  and $t_\text{long}^\text{p} = 14372 \ m^{-1}$, for waveguides separated by $g_\text{s}$ and $g_\text{l}$ respectively, were calculated using a finite difference eigenmode solver. The propagation of the pump forms a beating pattern between the two trivial modes, as demonstrated in Ref.~\cite{Blanco-Redondo2016} and illustrated here in Fig.~\ref{Dynamics}~(b).

As the pump propagates through the lattice, Fig.~\ref{Dynamics}~(b), SFWM generates biphoton states at signal and idler frequencies, as described in Eq.~\ref{eq:H-NL}. The photon pairs are generated at a rate determined by the local pump amplitude and by the nonlinear parameter $\gamma=120\ W^{-1}m^{-1}$. The generated signal and idler photons then propagate independently according to $H_\text{s,i}$, with coupling strengths $t_\text{short}^\text{s} = 43896 \ m^{-1}$, $t_\text{long}^\text{s} = 13892 \ m^{-1}$ for the signal photons, and $t_\text{short}^\text{i} = 46219 \ m^{-1}$ and $t_\text{long}^\text{i} = 14868 \ m^{-1}$ for the idler photons.

As shown in the biphoton probabilities in Fig.~\ref{Dynamics}~(c) very few biphotons are present initially, since the pump has not propagated for long enough to induce significant nonlinear interactions. However, as the pump propagates further along the lattice, the accumulated SFWM-generated biphotons increases monotonically. Even though the pump profile has no overlap with the topological mode, inter-modal SFWM allows photons to be created in both topological and trivial modes. As a result, the biphoton propagation exhibits a beating pattern involving the topological and the two trivial eigenmodes.

Figures~\ref{Dynamics}~(d-f) show the simulated biphoton probabilities of the quantum state at three different propagation points (A-C). It is evident that the entangled state takes vastly different spatial forms at different points, as expected from states at different stages of a beating pattern. Figures ~\ref{Dynamics}~(g-i) show the simulated biphoton population of the three relevant eigenmodes as a function of propagation distance, calculated by performing the overlap integral of the simulated biphoton probabilities at different propagation points with the amplitude of the relevant biphoton eigenmodes. Initially, signal and idler photons populate the trivial eigenmodes (Tr1, Tr2) almost exclusively. This is because the pump is fully localized in the center waveguide which has zero support for the topological mode. As the signal propagates, the beating pattern populates the waveguides adjacent to the center, and SFWM enables the generation of biphotons overlapping with the topological mode (Tp). Thus, the resulting biphotons are in a superposition of the topological and trivial states towards the middle and end propagation points. These matrices provide evidence that the weight of the topological mode in the entangled biphoton state grows with propagation length.

\paragraph{Measurements of the entangled biphoton states. --}

Using the experimental setup depicted in Fig.~\ref{ExpSketch}~(a) we measured the signal-idler correlation counts at the output of the five center waveguides over an interval of 120 seconds. We fabricated and measured lattices with three different levels of off-diagonal disorder, i.e. disorder in the coupling strengths between waveguides. This disorder is created by randomizing the position of each of the waveguides in our design. We quantify the disorder strength, $\delta \in [0,0.5]$, in terms of the proportion of variation of the coupling constants. For each fabricated lattice we know the engineered position of each waveguide. Therefore, we can accurately simulate the expected behaviour of the disordered lattices.

\begin{figure}
\centering
\includegraphics[width=\columnwidth]{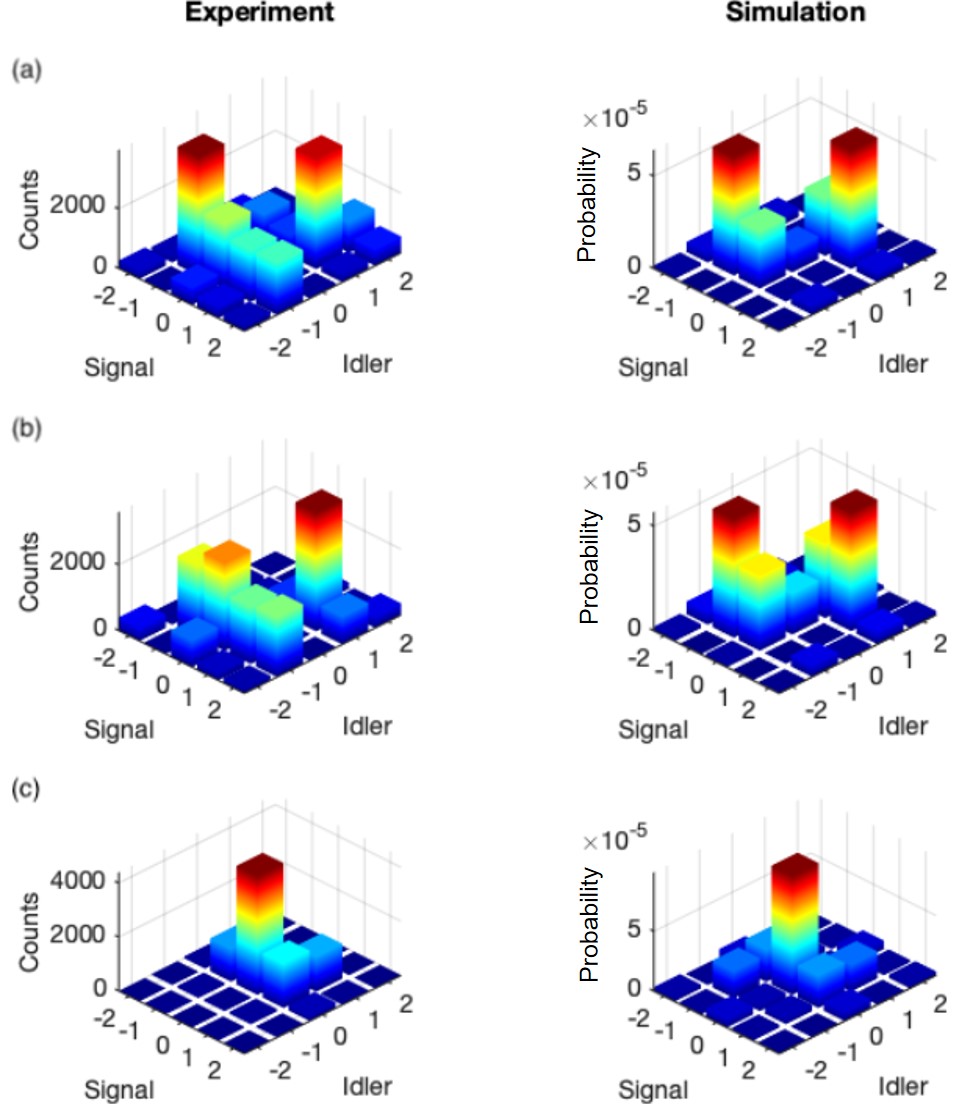}
\caption{Experimentally measured correlated photon counts (left) and simulated biphoton probabilities (right) at the lattice output with (a) $\delta=0$ (b) $\delta=0.2$ (c) $\delta=0.4$.}
\label{PhotonCounts}
\end{figure}

Figure~\ref{PhotonCounts}~(a) shows the experimentally measured (left) and simulated (right) spatial biphoton correlation at the output of the lattice with no introduced disorder $\delta = 0$, only the disorder inherent to nanofabrication, which for our platform, fabricated by a combination of electron beam lithography and reactive ion-etching, is expected to be at most 1\% of waveguide separation. The measured correlation map shows two maxima at positions [-1,-1] and [1,1], indicating that correlated signal and idler photons have a high probability of both being in waveguide -1 or in waveguide +1. There are also non-negligible correlation counts in other positions. Experiments and simulations match qualitatively with small differences due to nanofabrication imperfections. These results are consistent with the propagation simulations in Fig.~\ref{Dynamics}~(f), which show low photon counts in the center waveguide (0) and two maxima in the adjacent waveguides (-1 and 1) at the lattice output. Further, the Schmidt decomposition shows three coefficients clearly larger than the rest $S_{1-3} = [1, 0.22, 0.13]$, which is consistent with a three-mode entangled state, as predicted in Fig.~\ref{Dynamics}~(i). Note that the Schmidt decomposition is performed on the square root of the correlation counts. This calculation assumes a flat-phase distribution across the wavefunction, providing a lower-bound measurement of entanglement \cite{Harder:13}.

Next, we characterize the entanglement at the output of a lattice with moderate disorder, characterized by $\delta = 0.2$. The simulated biphoton correlation at the output, depicted in the right panel of Fig.~\ref{PhotonCounts}~(b), shows a similar correlation count distribution to the case without disorder: two maxima, and non-negligible counts in other waveguides. In the left panel measurements, however, one of the maxima has shifted from position [-1,-1] to [0, -1], highlighting that at this level of disorder, the entanglement starts to change for reasons that we analyze later. Nonetheless, we observe three Schmidt coefficients clearly larger than the rest, maintaining the same values as in the $\delta = 0$ case.

Finally, we explore a lattice with high disorder, $\delta = 0.4$. In this case, both measurements and simulations in Fig.~\ref{PhotonCounts}~(c) show a clear change in the output entangled state, with a maximum at the position [0,0]. In other words, signal and idler are both, with high probability, in the center waveguide. The Schmidt decomposition of the measured state at this disorder strength shows three coefficients larger than the rest, with two of these coefficients being significantly smaller than in the cases with lower disorder. This significant change tells us that the output biphoton correlation is not robust to high levels of disorder, which is understood as resulting from the contribution of the trivially localized modes in the entanglement, as we study in detail next.

\paragraph{Analysis of the robustness of the entangled biphoton states. --}

To investigate the effect of system disorder on the entangled biphoton states, we model our system using the above Hamiltonian and numerically characterize the relative weights of the components of the entangled states generated in the presence of disorder. We show how the eigenmode energy in the topological mode, as well as the support for the entangled biphoton state in the topological mode, remain robust to disorder, in contrast to the energy and support of the trivial modes.

\begin{figure}[b]
\centering
\includegraphics[width=1.05\columnwidth]{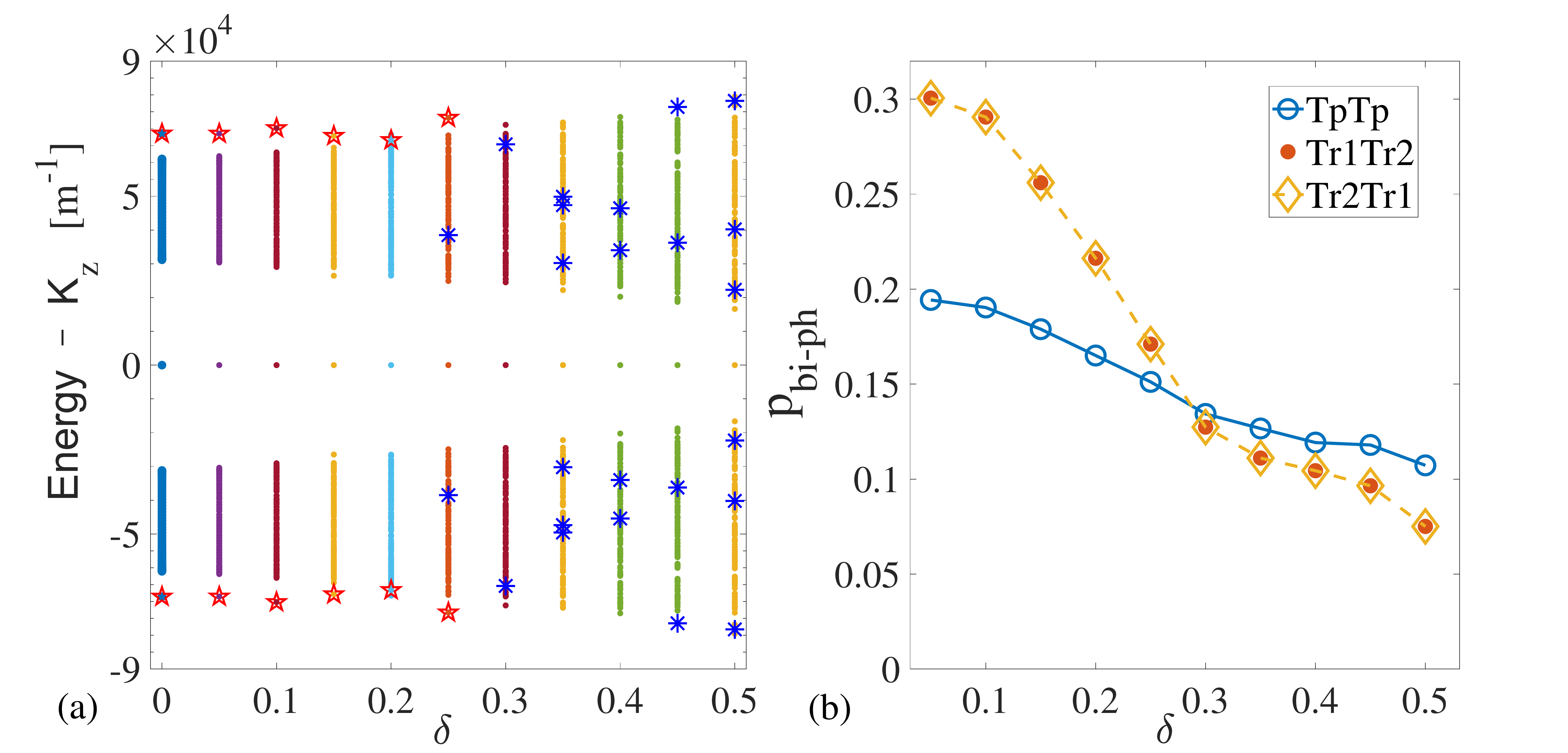}
\caption{(a) Relation between eigen-energy and disorder strength $\delta\in[0,0.5]$. The zero-energy topological mode is robust to disorder. The extended and topological eigenmodes at each level of disorder are plotted in different colors.
(b) Average relative weights as probabilities of topological and trivial biphoton eigenmodes, $p_{\text{bi{-}ph}}$,  with respect to the disorder strength.}
\label{Energy-Disorder}
\end{figure}

Figure~\ref{Energy-Disorder}~(a) shows the eigenmode energy distribution for different disorder strengths $\delta\in [0,0.5]$, corresponding to variations in the coupling constant from $0\%$ to $50\%$. At $\delta = 0$, we observe the presence of the topological mode in the center of the band gap, and the two trivial modes, also localized at the short-short defect, appearing at the lowest and highest energy of the diagram. These trivially localized modes at the energy extremes are marked with a red star. For increasing values of $\delta$, the energy of the topological mode remains pinned to zero-energy, all the way up to $\delta = 0.5$.

The energy of the trivially localized modes, however, changes significantly with disorder. From $\delta = 0.15$, the gap between the extended eigenmodes and the trivially localized eigenmodes begins to close, which leads to delocalization of these modes. From $\delta = 0.25$ we see the emergence of new trivially localized modes in the energy diagram, highlighted with blue asterisks. These modes are not localized in the original short-short defect but rather in other points of the lattice. Because of the nonlinear properties of the waveguides, these emergent eigenmodes also contribute to generating biphoton states. These modes emerge from the system disorder and their number generally increases with it. This can explain the changes observed experimentally on the entangled biphoton states generated in different lattices.

Next, we study the average relative weight of the topological and trivial components of the entangled state, $p_{\rm{ bi-ph}}$, for increasing levels of disorder in the range $\delta \in [0, 0.5]$.  Specifically, for each disorder strength we generated 400 different random configurations of disordered waveguide lattices and evaluated the average overlap between the generated biphoton states and the topological biphoton mode TpTp (blue empty dots in Fig.~\ref{Energy-Disorder}~(b)), as well as the hybrid trivial biphoton modes Tr1Tr2 (red solid dots) and Tr2Tr1 (empty yellow diamonds). At low levels of disorder, the population probability of the trivial biphoton modes is higher than that of the topological mode, with $30\%$ for both Tr1Tr2 and Tr2Tr1 and $19\%$ for TpTp. This is consistent with our predictions in Fig.~\ref{Dynamics}~(i): the population of the topological biphoton state increases with propagation distance, but, even at the output of the 381 $\mu m$-long lattice, it is still smaller than that of the trivial biphoton states.

The population probability of the trivial modes, however, decreases abruptly with increasing disorder, dropping from $30\%$ at $\delta = 0.05$ to a mere $7\%$ at $\delta = 0.5$. Consistently with what we observed in Fig.~\ref{Energy-Disorder}~(a), the abrupt decay commences at $\delta = 0.15$ where the gaps between the extended and the trivially localized eigenmodes start to close. This decay becomes more relaxed around $\delta = 0.25$, coinciding with the disorder strength at which we started to see the emergence of additional trivial states localized around the short-short defect.

The population probability of the topological biphoton mode decays at a slower rate, from $19\%$ at $\delta = 0.05$ to $11\%$ at $\delta = 0.5$. Indeed, for disorder strengths larger than $\delta = 0.3$, the population probabilty of the topological biphoton mode is higher than that of any of the hybrid trivial modes. This moderate decay in the population of the topological biphoton mode with disorder is related to the reduction of the size of the topological bandgap, as evidenced in Fig.~\ref{Energy-Disorder}~(a). Further, the decay softens around $\delta = 0.25$, just as the decay of the population of the hybrid trivial modes also softens. This suggests that, as we conjectured above, the emergence of the disorder-induced trivial states localized in the short-short defect contributes to the generation of biphotons in both the topological biphoton mode and the hybrid, trivially localized biphoton modes.

%\paragraph{Conclusions.}

In conclusion, we have characterized the robustness of the components of entangled states across spatial modes of different topological character. The experimental measurements of the correlated biphoton counts, combined with simulations and eigenmode analysis, evidence the entanglement between modes of distinct topological nature. The measured entangled biphoton state does not preserve its shape under the presence of off-diagonal disorder, which is explained by the large weight of the trivial biphoton modes in such entanglement, and the fact that, while the underlying topological eigenmode is robust to off-diagonal disorder the trivial eigenmodes are not.  Our results can, however, be generalized to systems that can support entanglement of co-localized topologically-distinct modes, all of them topologically protected, such as multiband photonic superlattices \cite{PhysRevA.98.043838, PhysRevLett.126.066401} and multimode one-way waveguides of large Chern numbers \cite{Skirlo2014}.

Finally, we have shown how the topological mode can be excited indirectly, via interaction with trivial modes that are easier to excite. This photonic analogue could be leveraged by the quantum information community to gain insights into how to excite topologically protected modes, such as Majorana modes, that hold enormous promise for quantum information technologies but whose protected nature makes them hard to excite. Furthermore, the promise of entanglement between topological and non-topological modes gives a potential mechanism to further manipulate information in the topological mode indirectly, suggesting avenues for topological quantum information transfer \cite{PhysRevB.103.085129}.

\begin{acknowledgements}
We acknowledge support from the Australian Research Council (ARC) via the Centre of Excellence in Engineered Quantum Systems (EQuS) project number CE170100009. BB is supported by a European Commission Marie Skłodowska Curie Individual Fellowship (FrEQuMP, 846073). WZ appreciates financial support from NSFC (Grant No.  12104101).

C.D. and W.-W.Z. contributed equally to this work.
\end{acknowledgements}
\bibliographystyle{apsrev4-1}
\bibliography{apssamp}% Produces the bibliography via BibTeX.

\providecommand{\noopsort}[1]{}\providecommand{\singleletter}[1]{#1}%
\begin{thebibliography}{29}%
\makeatletter
\providecommand \@ifxundefined [1]{%
 \@ifx{#1\undefined}
}%
\providecommand \@ifnum [1]{%
 \ifnum #1\expandafter \@firstoftwo
 \else \expandafter \@secondoftwo
 \fi
}%
\providecommand \@ifx [1]{%
 \ifx #1\expandafter \@firstoftwo
 \else \expandafter \@secondoftwo
 \fi
}%
\providecommand \natexlab [1]{#1}%
\providecommand \enquote  [1]{``#1''}%
\providecommand \bibnamefont  [1]{#1}%
\providecommand \bibfnamefont [1]{#1}%
\providecommand \citenamefont [1]{#1}%
\providecommand \href@noop [0]{\@secondoftwo}%
\providecommand \href [0]{\begingroup \@sanitize@url \@href}%
\providecommand \@href[1]{\@@startlink{#1}\@@href}%
\providecommand \@@href[1]{\endgroup#1\@@endlink}%
\providecommand \@sanitize@url [0]{\catcode `\\12\catcode `\$12\catcode
  `\&12\catcode `\#12\catcode `\^12\catcode `\_12\catcode `\%12\relax}%
\providecommand \@@startlink[1]{}%
\providecommand \@@endlink[0]{}%
\providecommand \url  [0]{\begingroup\@sanitize@url \@url }%
\providecommand \@url [1]{\endgroup\@href {#1}{\urlprefix }}%
\providecommand \urlprefix  [0]{URL }%
\providecommand \Eprint [0]{\href }%
\providecommand \doibase [0]{http://dx.doi.org/}%
\providecommand \selectlanguage [0]{\@gobble}%
\providecommand \bibinfo  [0]{\@secondoftwo}%
\providecommand \bibfield  [0]{\@secondoftwo}%
\providecommand \translation [1]{[#1]}%
\providecommand \BibitemOpen [0]{}%
\providecommand \bibitemStop [0]{}%
\providecommand \bibitemNoStop [0]{.\EOS\space}%
\providecommand \EOS [0]{\spacefactor3000\relax}%
\providecommand \BibitemShut  [1]{\csname bibitem#1\endcsname}%
\let\auto@bib@innerbib\@empty
%</preamble>
\bibitem [{\citenamefont {Klitzing}\ \emph {et~al.}(1980)\citenamefont
  {Klitzing}, \citenamefont {Dorda},\ and\ \citenamefont
  {Pepper}}]{Klitzing1980}%
  \BibitemOpen
  \bibfield  {author} {\bibinfo {author} {\bibfnamefont {K.~V.}\ \bibnamefont
  {Klitzing}}, \bibinfo {author} {\bibfnamefont {G.}~\bibnamefont {Dorda}}, \
  and\ \bibinfo {author} {\bibfnamefont {M.}~\bibnamefont {Pepper}},\ }\href
  {\doibase 10.1103/PhysRevLett.45.494} {\bibfield  {journal} {\bibinfo
  {journal} {Physical Review Letters}\ }\textbf {\bibinfo {volume} {45}},\
  \bibinfo {pages} {494} (\bibinfo {year} {1980})},\ \Eprint
  {http://arxiv.org/abs/arXiv:1011.1669v3} {arXiv:arXiv:1011.1669v3}
  \BibitemShut {NoStop}%
\bibitem [{\citenamefont {Thouless}\ \emph {et~al.}(1982)\citenamefont
  {Thouless}, \citenamefont {Kohmoto}, \citenamefont {Nightingale},\ and\
  \citenamefont {{Den Nijs}}}]{Thouless1982}%
  \BibitemOpen
  \bibfield  {author} {\bibinfo {author} {\bibfnamefont {D.~J.}\ \bibnamefont
  {Thouless}}, \bibinfo {author} {\bibfnamefont {M.}~\bibnamefont {Kohmoto}},
  \bibinfo {author} {\bibfnamefont {M.~P.}\ \bibnamefont {Nightingale}}, \ and\
  \bibinfo {author} {\bibfnamefont {M.}~\bibnamefont {{Den Nijs}}},\ }\href
  {\doibase 10.1103/PhysRevLett.49.405} {\bibfield  {journal} {\bibinfo
  {journal} {Physical Review Letters}\ }\textbf {\bibinfo {volume} {49}},\
  \bibinfo {pages} {405} (\bibinfo {year} {1982})},\ \Eprint
  {http://arxiv.org/abs/arXiv:1011.1669v3} {arXiv:arXiv:1011.1669v3}
  \BibitemShut {NoStop}%
\bibitem [{\citenamefont {Ozawa}\ \emph {et~al.}(2019)\citenamefont {Ozawa},
  \citenamefont {Price}, \citenamefont {Amo}, \citenamefont {Goldman},
  \citenamefont {Hafezi}, \citenamefont {Lu}, \citenamefont {Rechtsman},
  \citenamefont {Schuster}, \citenamefont {Simon}, \citenamefont {Zilberberg},\
  and\ \citenamefont {Carusotto}}]{Ozawa2019}%
  \BibitemOpen
  \bibfield  {author} {\bibinfo {author} {\bibfnamefont {T.}~\bibnamefont
  {Ozawa}}, \bibinfo {author} {\bibfnamefont {H.~M.}\ \bibnamefont {Price}},
  \bibinfo {author} {\bibfnamefont {A.}~\bibnamefont {Amo}}, \bibinfo {author}
  {\bibfnamefont {N.}~\bibnamefont {Goldman}}, \bibinfo {author} {\bibfnamefont
  {M.}~\bibnamefont {Hafezi}}, \bibinfo {author} {\bibfnamefont
  {L.}~\bibnamefont {Lu}}, \bibinfo {author} {\bibfnamefont {M.~C.}\
  \bibnamefont {Rechtsman}}, \bibinfo {author} {\bibfnamefont {D.}~\bibnamefont
  {Schuster}}, \bibinfo {author} {\bibfnamefont {J.}~\bibnamefont {Simon}},
  \bibinfo {author} {\bibfnamefont {O.}~\bibnamefont {Zilberberg}}, \ and\
  \bibinfo {author} {\bibfnamefont {I.}~\bibnamefont {Carusotto}},\ }\href@noop
  {} {\bibfield  {journal} {\bibinfo  {journal} {Reviews of Modern Physics}\
  }\textbf {\bibinfo {volume} {91}},\ \bibinfo {pages} {015006} (\bibinfo
  {year} {2019})}\BibitemShut {NoStop}%
\bibitem [{\citenamefont {Ryu}\ and\ \citenamefont {Hatsugai}(2002)}]{Ryu2002}%
  \BibitemOpen
  \bibfield  {author} {\bibinfo {author} {\bibfnamefont {S.}~\bibnamefont
  {Ryu}}\ and\ \bibinfo {author} {\bibfnamefont {Y.}~\bibnamefont {Hatsugai}},\
  }\href {\doibase 10.1103/PhysRevLett.89.077002} {\bibfield  {journal}
  {\bibinfo  {journal} {Physical Review Letters}\ }\textbf {\bibinfo {volume}
  {89}},\ \bibinfo {pages} {077002} (\bibinfo {year} {2002})}\BibitemShut
  {NoStop}%
\bibitem [{\citenamefont {Wang}\ \emph {et~al.}(2009)\citenamefont {Wang},
  \citenamefont {Chong}, \citenamefont {Joannopoulos},\ and\ \citenamefont
  {Solja{\v{c}}i{\'{c}}}}]{Wang2009}%
  \BibitemOpen
  \bibfield  {author} {\bibinfo {author} {\bibfnamefont {Z.}~\bibnamefont
  {Wang}}, \bibinfo {author} {\bibfnamefont {Y.}~\bibnamefont {Chong}},
  \bibinfo {author} {\bibfnamefont {J.~D.}\ \bibnamefont {Joannopoulos}}, \
  and\ \bibinfo {author} {\bibfnamefont {M.}~\bibnamefont
  {Solja{\v{c}}i{\'{c}}}},\ }\href {\doibase 10.1038/nature08293} {\bibfield
  {journal} {\bibinfo  {journal} {Nature}\ }\textbf {\bibinfo {volume} {461}},\
  \bibinfo {pages} {772} (\bibinfo {year} {2009})},\ \Eprint
  {http://arxiv.org/abs/1507.03002} {arXiv:1507.03002} \BibitemShut {NoStop}%
\bibitem [{\citenamefont {Kraus}\ \emph {et~al.}(2012)\citenamefont {Kraus},
  \citenamefont {Lahini}, \citenamefont {Ringel}, \citenamefont {Verbin},\ and\
  \citenamefont {Zilberberg}}]{Kraus2012}%
  \BibitemOpen
  \bibfield  {author} {\bibinfo {author} {\bibfnamefont {Y.~E.}\ \bibnamefont
  {Kraus}}, \bibinfo {author} {\bibfnamefont {Y.}~\bibnamefont {Lahini}},
  \bibinfo {author} {\bibfnamefont {Z.}~\bibnamefont {Ringel}}, \bibinfo
  {author} {\bibfnamefont {M.}~\bibnamefont {Verbin}}, \ and\ \bibinfo {author}
  {\bibfnamefont {O.}~\bibnamefont {Zilberberg}},\ }\href {\doibase
  10.1103/PhysRevLett.109.106402} {\bibfield  {journal} {\bibinfo  {journal}
  {Physical Review Letters}\ }\textbf {\bibinfo {volume} {109}},\ \bibinfo
  {pages} {106402} (\bibinfo {year} {2012})}\BibitemShut {NoStop}%
\bibitem [{\citenamefont {Rechtsman}\ \emph {et~al.}(2013)\citenamefont
  {Rechtsman}, \citenamefont {Zeuner}, \citenamefont {Plotnik}, \citenamefont
  {Lumer}, \citenamefont {Podolsky}, \citenamefont {Dreisow}, \citenamefont
  {Nolte}, \citenamefont {Segev},\ and\ \citenamefont
  {Szameit}}]{Rechtsman2014}%
  \BibitemOpen
  \bibfield  {author} {\bibinfo {author} {\bibfnamefont {M.~C.}\ \bibnamefont
  {Rechtsman}}, \bibinfo {author} {\bibfnamefont {J.~M.}\ \bibnamefont
  {Zeuner}}, \bibinfo {author} {\bibfnamefont {Y.}~\bibnamefont {Plotnik}},
  \bibinfo {author} {\bibfnamefont {Y.}~\bibnamefont {Lumer}}, \bibinfo
  {author} {\bibfnamefont {D.}~\bibnamefont {Podolsky}}, \bibinfo {author}
  {\bibfnamefont {F.}~\bibnamefont {Dreisow}}, \bibinfo {author} {\bibfnamefont
  {S.}~\bibnamefont {Nolte}}, \bibinfo {author} {\bibfnamefont
  {M.}~\bibnamefont {Segev}}, \ and\ \bibinfo {author} {\bibfnamefont
  {A.}~\bibnamefont {Szameit}},\ }\href {\doibase 10.1038/nature12066}
  {\bibfield  {journal} {\bibinfo  {journal} {Nature}\ }\textbf {\bibinfo
  {volume} {496}},\ \bibinfo {pages} {196} (\bibinfo {year}
  {2013})}\BibitemShut {NoStop}%
\bibitem [{\citenamefont {Hafezi}\ \emph {et~al.}(2013)\citenamefont {Hafezi},
  \citenamefont {Mittal}, \citenamefont {Fan}, \citenamefont {Migdall},\ and\
  \citenamefont {Taylor}}]{Hafezi2013}%
  \BibitemOpen
  \bibfield  {author} {\bibinfo {author} {\bibfnamefont {M.}~\bibnamefont
  {Hafezi}}, \bibinfo {author} {\bibfnamefont {S.}~\bibnamefont {Mittal}},
  \bibinfo {author} {\bibfnamefont {J.}~\bibnamefont {Fan}}, \bibinfo {author}
  {\bibfnamefont {A.}~\bibnamefont {Migdall}}, \ and\ \bibinfo {author}
  {\bibfnamefont {J.~M.}\ \bibnamefont {Taylor}},\ }\href {\doibase
  10.1038/nphoton.2013.274} {\bibfield  {journal} {\bibinfo  {journal} {Nature
  Photonics}\ }\textbf {\bibinfo {volume} {7}},\ \bibinfo {pages} {1001}
  (\bibinfo {year} {2013})}\BibitemShut {NoStop}%
\bibitem [{\citenamefont {Blanco-Redondo}(2020)}]{Blanco-Redondo2020}%
  \BibitemOpen
  \bibfield  {author} {\bibinfo {author} {\bibfnamefont {A.}~\bibnamefont
  {Blanco-Redondo}},\ }\href {\doibase 10.1109/JPROC.2019.2939987} {\bibfield
  {journal} {\bibinfo  {journal} {Proceedings of the IEEE}\ }\textbf {\bibinfo
  {volume} {108}},\ \bibinfo {pages} {837} (\bibinfo {year}
  {2020})}\BibitemShut {NoStop}%
\bibitem [{\citenamefont {Barik}\ \emph {et~al.}(2018)\citenamefont {Barik},
  \citenamefont {Karasahin}, \citenamefont {Flower}, \citenamefont {Cai},
  \citenamefont {Miyake}, \citenamefont {DeGottardi}, \citenamefont {Hafezi},\
  and\ \citenamefont {Waks}}]{Barik2018}%
  \BibitemOpen
  \bibfield  {author} {\bibinfo {author} {\bibfnamefont {S.}~\bibnamefont
  {Barik}}, \bibinfo {author} {\bibfnamefont {A.}~\bibnamefont {Karasahin}},
  \bibinfo {author} {\bibfnamefont {C.}~\bibnamefont {Flower}}, \bibinfo
  {author} {\bibfnamefont {T.}~\bibnamefont {Cai}}, \bibinfo {author}
  {\bibfnamefont {H.}~\bibnamefont {Miyake}}, \bibinfo {author} {\bibfnamefont
  {W.}~\bibnamefont {DeGottardi}}, \bibinfo {author} {\bibfnamefont
  {M.}~\bibnamefont {Hafezi}}, \ and\ \bibinfo {author} {\bibfnamefont
  {E.}~\bibnamefont {Waks}},\ }\href {\doibase 10.1126/science.aaq0327}
  {\bibfield  {journal} {\bibinfo  {journal} {Science}\ }\textbf {\bibinfo
  {volume} {359}},\ \bibinfo {pages} {666} (\bibinfo {year}
  {2018})}\BibitemShut {NoStop}%
\bibitem [{\citenamefont {Wang}\ \emph
  {et~al.}(2019{\natexlab{a}})\citenamefont {Wang}, \citenamefont {Lu},
  \citenamefont {Mei}, \citenamefont {Gao}, \citenamefont {Li}, \citenamefont
  {Tang}, \citenamefont {Zhu}, \citenamefont {Jia},\ and\ \citenamefont
  {Jin}}]{Wang2019b}%
  \BibitemOpen
  \bibfield  {author} {\bibinfo {author} {\bibfnamefont {Y.}~\bibnamefont
  {Wang}}, \bibinfo {author} {\bibfnamefont {Y.~H.}\ \bibnamefont {Lu}},
  \bibinfo {author} {\bibfnamefont {F.}~\bibnamefont {Mei}}, \bibinfo {author}
  {\bibfnamefont {J.}~\bibnamefont {Gao}}, \bibinfo {author} {\bibfnamefont
  {Z.~M.}\ \bibnamefont {Li}}, \bibinfo {author} {\bibfnamefont
  {H.}~\bibnamefont {Tang}}, \bibinfo {author} {\bibfnamefont {S.~L.}\
  \bibnamefont {Zhu}}, \bibinfo {author} {\bibfnamefont {S.}~\bibnamefont
  {Jia}}, \ and\ \bibinfo {author} {\bibfnamefont {X.~M.}\ \bibnamefont
  {Jin}},\ }\href {\doibase 10.1103/PhysRevLett.122.193903} {\bibfield
  {journal} {\bibinfo  {journal} {Physical Review Letters}\ }\textbf {\bibinfo
  {volume} {122}},\ \bibinfo {pages} {193903} (\bibinfo {year}
  {2019}{\natexlab{a}})}\BibitemShut {NoStop}%
\bibitem [{\citenamefont {Mittal}\ \emph {et~al.}(2018)\citenamefont {Mittal},
  \citenamefont {Goldschmidt},\ and\ \citenamefont {Hafezi}}]{Mittal2018}%
  \BibitemOpen
  \bibfield  {author} {\bibinfo {author} {\bibfnamefont {S.}~\bibnamefont
  {Mittal}}, \bibinfo {author} {\bibfnamefont {E.~A.}\ \bibnamefont
  {Goldschmidt}}, \ and\ \bibinfo {author} {\bibfnamefont {M.}~\bibnamefont
  {Hafezi}},\ }\href {\doibase 10.1038/s41586-018-0478-3} {\bibfield  {journal}
  {\bibinfo  {journal} {Nature}\ }\textbf {\bibinfo {volume} {561}},\ \bibinfo
  {pages} {502} (\bibinfo {year} {2018})}\BibitemShut {NoStop}%
\bibitem [{\citenamefont {Blanco-Redondo}\ \emph {et~al.}(2018)\citenamefont
  {Blanco-Redondo}, \citenamefont {Bell}, \citenamefont {Oren}, \citenamefont
  {Eggleton},\ and\ \citenamefont {Segev}}]{Blanco-Redondo2018}%
  \BibitemOpen
  \bibfield  {author} {\bibinfo {author} {\bibfnamefont {A.}~\bibnamefont
  {Blanco-Redondo}}, \bibinfo {author} {\bibfnamefont {B.}~\bibnamefont
  {Bell}}, \bibinfo {author} {\bibfnamefont {D.}~\bibnamefont {Oren}}, \bibinfo
  {author} {\bibfnamefont {B.~J.}\ \bibnamefont {Eggleton}}, \ and\ \bibinfo
  {author} {\bibfnamefont {M.}~\bibnamefont {Segev}},\ }\href {\doibase
  10.1126/science.aau4296} {\bibfield  {journal} {\bibinfo  {journal}
  {Science}\ }\textbf {\bibinfo {volume} {362}},\ \bibinfo {pages} {568}
  (\bibinfo {year} {2018})},\ \Eprint
  {http://arxiv.org/abs/https://science.sciencemag.org/content/362/6414/568.full.pdf}
  {https://science.sciencemag.org/content/362/6414/568.full.pdf} \BibitemShut
  {NoStop}%
\bibitem [{\citenamefont {Wang}\ \emph
  {et~al.}(2019{\natexlab{b}})\citenamefont {Wang}, \citenamefont {Doyle},
  \citenamefont {Bell}, \citenamefont {Collins}, \citenamefont {Magi},
  \citenamefont {Eggleton}, \citenamefont {Segev},\ and\ \citenamefont
  {Blanco-Redondo}}]{Wang2019}%
  \BibitemOpen
  \bibfield  {author} {\bibinfo {author} {\bibfnamefont {M.}~\bibnamefont
  {Wang}}, \bibinfo {author} {\bibfnamefont {C.}~\bibnamefont {Doyle}},
  \bibinfo {author} {\bibfnamefont {B.}~\bibnamefont {Bell}}, \bibinfo {author}
  {\bibfnamefont {M.~J.}\ \bibnamefont {Collins}}, \bibinfo {author}
  {\bibfnamefont {E.}~\bibnamefont {Magi}}, \bibinfo {author} {\bibfnamefont
  {B.~J.}\ \bibnamefont {Eggleton}}, \bibinfo {author} {\bibfnamefont
  {M.}~\bibnamefont {Segev}}, \ and\ \bibinfo {author} {\bibfnamefont
  {A.}~\bibnamefont {Blanco-Redondo}},\ }\href {\doibase
  10.1515/nanoph-2019-0058} {\bibfield  {journal} {\bibinfo  {journal}
  {Nanophotonics}\ }\textbf {\bibinfo {volume} {8}},\ \bibinfo {pages} {1327}
  (\bibinfo {year} {2019}{\natexlab{b}})}\BibitemShut {NoStop}%
\bibitem [{\citenamefont {Wang}\ \emph {et~al.}(shed)\citenamefont {Wang},
  \citenamefont {Lu}, \citenamefont {Gao}, \citenamefont {Ren}, \citenamefont
  {Chang}, \citenamefont {Jiao}, \citenamefont {Zhang},\ and\ \citenamefont
  {Jin}}]{wang2019topologically}%
  \BibitemOpen
  \bibfield  {author} {\bibinfo {author} {\bibfnamefont {Y.}~\bibnamefont
  {Wang}}, \bibinfo {author} {\bibfnamefont {Y.-H.}\ \bibnamefont {Lu}},
  \bibinfo {author} {\bibfnamefont {J.}~\bibnamefont {Gao}}, \bibinfo {author}
  {\bibfnamefont {R.-J.}\ \bibnamefont {Ren}}, \bibinfo {author} {\bibfnamefont
  {Y.-J.}\ \bibnamefont {Chang}}, \bibinfo {author} {\bibfnamefont {Z.-Q.}\
  \bibnamefont {Jiao}}, \bibinfo {author} {\bibfnamefont {Z.-Y.}\ \bibnamefont
  {Zhang}}, \ and\ \bibinfo {author} {\bibfnamefont {X.-M.}\ \bibnamefont
  {Jin}},\ }\href@noop {} {\bibfield  {journal} {\bibinfo  {journal} {arXiv}\ }
  (\bibinfo {year} {unpublished})},\ \Eprint {http://arxiv.org/abs/1903.03015}
  {arXiv:1903.03015 [quant-ph]} \BibitemShut {NoStop}%
\bibitem [{\citenamefont {Tambasco}\ \emph {et~al.}(2018)\citenamefont
  {Tambasco}, \citenamefont {Corrielli}, \citenamefont {Chapman}, \citenamefont
  {Crespi}, \citenamefont {Zilberberg}, \citenamefont {Osellame},\ and\
  \citenamefont {Peruzzo}}]{Tombasco2018}%
  \BibitemOpen
  \bibfield  {author} {\bibinfo {author} {\bibfnamefont {J.~L.}\ \bibnamefont
  {Tambasco}}, \bibinfo {author} {\bibfnamefont {G.}~\bibnamefont {Corrielli}},
  \bibinfo {author} {\bibfnamefont {R.~J.}\ \bibnamefont {Chapman}}, \bibinfo
  {author} {\bibfnamefont {A.}~\bibnamefont {Crespi}}, \bibinfo {author}
  {\bibfnamefont {O.}~\bibnamefont {Zilberberg}}, \bibinfo {author}
  {\bibfnamefont {R.}~\bibnamefont {Osellame}}, \ and\ \bibinfo {author}
  {\bibfnamefont {A.}~\bibnamefont {Peruzzo}},\ }\href {\doibase
  10.1126/sciadv.aat3187} {\bibfield  {journal} {\bibinfo  {journal} {Science
  Advances}\ }\textbf {\bibinfo {volume} {4}},\ \bibinfo {pages} {1} (\bibinfo
  {year} {2018})}\BibitemShut {NoStop}%
\bibitem [{\citenamefont {Gorlach}\ and\ \citenamefont
  {Poddubny}(2017)}]{Gorlach2017}%
  \BibitemOpen
  \bibfield  {author} {\bibinfo {author} {\bibfnamefont {M.~A.}\ \bibnamefont
  {Gorlach}}\ and\ \bibinfo {author} {\bibfnamefont {A.~N.}\ \bibnamefont
  {Poddubny}},\ }\href {\doibase 10.1103/PhysRevA.95.033831} {\bibfield
  {journal} {\bibinfo  {journal} {Physical Review A}\ }\textbf {\bibinfo
  {volume} {95}},\ \bibinfo {pages} {033831} (\bibinfo {year}
  {2017})}\BibitemShut {NoStop}%
\bibitem [{\citenamefont {Stepanenko}\ and\ \citenamefont
  {Gorlach}(2020)}]{Stepanenko2020}%
  \BibitemOpen
  \bibfield  {author} {\bibinfo {author} {\bibfnamefont {A.~A.}\ \bibnamefont
  {Stepanenko}}\ and\ \bibinfo {author} {\bibfnamefont {M.~A.}\ \bibnamefont
  {Gorlach}},\ }\href {\doibase 10.1103/PhysRevA.102.013510} {\bibfield
  {journal} {\bibinfo  {journal} {Physical Review A}\ }\textbf {\bibinfo
  {volume} {102}},\ \bibinfo {pages} {013510} (\bibinfo {year}
  {2020})}\BibitemShut {NoStop}%
\bibitem [{\citenamefont {Olekhno}\ \emph {et~al.}(2020)\citenamefont
  {Olekhno}, \citenamefont {Kretov}, \citenamefont {Stepanenko}, \citenamefont
  {Ivanova}, \citenamefont {Yaroshenko}, \citenamefont {Puhtina}, \citenamefont
  {Filonov}, \citenamefont {Cappello}, \citenamefont {Matekovits},\ and\
  \citenamefont {Gorlach}}]{Olekhno2020}%
  \BibitemOpen
  \bibfield  {author} {\bibinfo {author} {\bibfnamefont {N.~A.}\ \bibnamefont
  {Olekhno}}, \bibinfo {author} {\bibfnamefont {E.~I.}\ \bibnamefont {Kretov}},
  \bibinfo {author} {\bibfnamefont {A.~A.}\ \bibnamefont {Stepanenko}},
  \bibinfo {author} {\bibfnamefont {P.~A.}\ \bibnamefont {Ivanova}}, \bibinfo
  {author} {\bibfnamefont {V.~V.}\ \bibnamefont {Yaroshenko}}, \bibinfo
  {author} {\bibfnamefont {E.~M.}\ \bibnamefont {Puhtina}}, \bibinfo {author}
  {\bibfnamefont {D.~S.}\ \bibnamefont {Filonov}}, \bibinfo {author}
  {\bibfnamefont {B.}~\bibnamefont {Cappello}}, \bibinfo {author}
  {\bibfnamefont {L.}~\bibnamefont {Matekovits}}, \ and\ \bibinfo {author}
  {\bibfnamefont {M.~A.}\ \bibnamefont {Gorlach}},\ }\href {\doibase
  10.1038/s41467-020-14994-7} {\bibfield  {journal} {\bibinfo  {journal}
  {Nature Communications}\ }\textbf {\bibinfo {volume} {11}},\ \bibinfo {pages}
  {1436} (\bibinfo {year} {2020})}\BibitemShut {NoStop}%
\bibitem [{\citenamefont {Bergamasco}\ and\ \citenamefont
  {Liscidini}(2019)}]{Bergamasco2019}%
  \BibitemOpen
  \bibfield  {author} {\bibinfo {author} {\bibfnamefont {N.}~\bibnamefont
  {Bergamasco}}\ and\ \bibinfo {author} {\bibfnamefont {M.}~\bibnamefont
  {Liscidini}},\ }\href {\doibase 10.1103/PhysRevA.100.053827} {\bibfield
  {journal} {\bibinfo  {journal} {Physical Review A}\ }\textbf {\bibinfo
  {volume} {100}},\ \bibinfo {pages} {053827} (\bibinfo {year}
  {2019})}\BibitemShut {NoStop}%
\bibitem [{\citenamefont {Tschernig}\ \emph {et~al.}(2021)\citenamefont
  {Tschernig}, \citenamefont {Jimenez-Galan}, \citenamefont {Christodoulides},
  \citenamefont {Ivanov}, \citenamefont {Busch}, \citenamefont {Bandres},\ and\
  \citenamefont {Perez-Leija}}]{Tschernig2020}%
  \BibitemOpen
  \bibfield  {author} {\bibinfo {author} {\bibfnamefont {K.}~\bibnamefont
  {Tschernig}}, \bibinfo {author} {\bibfnamefont {A.}~\bibnamefont
  {Jimenez-Galan}}, \bibinfo {author} {\bibfnamefont {D.~N.}\ \bibnamefont
  {Christodoulides}}, \bibinfo {author} {\bibfnamefont {M.}~\bibnamefont
  {Ivanov}}, \bibinfo {author} {\bibfnamefont {K.}~\bibnamefont {Busch}},
  \bibinfo {author} {\bibfnamefont {M.~A.}\ \bibnamefont {Bandres}}, \ and\
  \bibinfo {author} {\bibfnamefont {A.}~\bibnamefont {Perez-Leija}},\
  }\href@noop {} {\bibfield  {journal} {\bibinfo  {journal} {Nature
  Communications}\ }\textbf {\bibinfo {volume} {12}},\ \bibinfo {pages} {1974}
  (\bibinfo {year} {2021})}\BibitemShut {NoStop}%
\bibitem [{\citenamefont {Zhang}\ \emph {et~al.}(2016)\citenamefont {Zhang},
  \citenamefont {Ding}, \citenamefont {Dong}, \citenamefont {Shi},
  \citenamefont {Wang}, \citenamefont {Liu}, \citenamefont {Li}, \citenamefont
  {Zhou}, \citenamefont {Shi},\ and\ \citenamefont {Guo}}]{MultipleDOF}%
  \BibitemOpen
  \bibfield  {author} {\bibinfo {author} {\bibfnamefont {W.}~\bibnamefont
  {Zhang}}, \bibinfo {author} {\bibfnamefont {D.}~\bibnamefont {Ding}},
  \bibinfo {author} {\bibfnamefont {M.-X.}\ \bibnamefont {Dong}}, \bibinfo
  {author} {\bibfnamefont {S.}~\bibnamefont {Shi}}, \bibinfo {author}
  {\bibfnamefont {K.}~\bibnamefont {Wang}}, \bibinfo {author} {\bibfnamefont
  {S.-L.}\ \bibnamefont {Liu}}, \bibinfo {author} {\bibfnamefont
  {Y.}~\bibnamefont {Li}}, \bibinfo {author} {\bibfnamefont {Z.-Y.}\
  \bibnamefont {Zhou}}, \bibinfo {author} {\bibfnamefont {B.-S.}\ \bibnamefont
  {Shi}}, \ and\ \bibinfo {author} {\bibfnamefont {G.-C.}\ \bibnamefont
  {Guo}},\ }\href {https://doi.org/10.1038/ncomms13514} {\bibfield  {journal}
  {\bibinfo  {journal} {Nat. Commun.}\ }\textbf {\bibinfo {volume} {7}},\
  \bibinfo {pages} {13514} (\bibinfo {year} {2016})}\BibitemShut {NoStop}%
\bibitem [{\citenamefont {Blanco-Redondo}\ \emph {et~al.}(2016)\citenamefont
  {Blanco-Redondo}, \citenamefont {Andonegui}, \citenamefont {Collins},
  \citenamefont {Harari}, \citenamefont {Lumer}, \citenamefont {Rechtsman},
  \citenamefont {Eggleton},\ and\ \citenamefont {Segev}}]{Blanco-Redondo2016}%
  \BibitemOpen
  \bibfield  {author} {\bibinfo {author} {\bibfnamefont {A.}~\bibnamefont
  {Blanco-Redondo}}, \bibinfo {author} {\bibfnamefont {I.}~\bibnamefont
  {Andonegui}}, \bibinfo {author} {\bibfnamefont {M.~J.}\ \bibnamefont
  {Collins}}, \bibinfo {author} {\bibfnamefont {G.}~\bibnamefont {Harari}},
  \bibinfo {author} {\bibfnamefont {Y.}~\bibnamefont {Lumer}}, \bibinfo
  {author} {\bibfnamefont {M.~C.}\ \bibnamefont {Rechtsman}}, \bibinfo {author}
  {\bibfnamefont {B.~J.}\ \bibnamefont {Eggleton}}, \ and\ \bibinfo {author}
  {\bibfnamefont {M.}~\bibnamefont {Segev}},\ }\href {\doibase
  10.1103/PhysRevLett.116.163901} {\bibfield  {journal} {\bibinfo  {journal}
  {Physical Review Letters}\ }\textbf {\bibinfo {volume} {116}},\ \bibinfo
  {pages} {163901} (\bibinfo {year} {2016})}\BibitemShut {NoStop}%
\bibitem [{\citenamefont {Su}\ \emph {et~al.}(1979)\citenamefont {Su},
  \citenamefont {Schrieffer},\ and\ \citenamefont {Heeger}}]{Su1979}%
  \BibitemOpen
  \bibfield  {author} {\bibinfo {author} {\bibfnamefont {W.~P.}\ \bibnamefont
  {Su}}, \bibinfo {author} {\bibfnamefont {J.~R.}\ \bibnamefont {Schrieffer}},
  \ and\ \bibinfo {author} {\bibfnamefont {A.~J.}\ \bibnamefont {Heeger}},\
  }\href {\doibase 10.1103/PhysRevLett.42.1698} {\bibfield  {journal} {\bibinfo
   {journal} {Physical Review Letters}\ }\textbf {\bibinfo {volume} {42}},\
  \bibinfo {pages} {1698} (\bibinfo {year} {1979})},\ \Eprint
  {http://arxiv.org/abs/PhysRevLett.42.1698} {arXiv:PhysRevLett.42.1698
  [10.1103]} \BibitemShut {NoStop}%
\bibitem [{\citenamefont {Harder}\ \emph {et~al.}(2013)\citenamefont {Harder},
  \citenamefont {Ansari}, \citenamefont {Brecht}, \citenamefont {Dirmeier},
  \citenamefont {Marquardt},\ and\ \citenamefont {Silberhorn}}]{Harder:13}%
  \BibitemOpen
  \bibfield  {author} {\bibinfo {author} {\bibfnamefont {G.}~\bibnamefont
  {Harder}}, \bibinfo {author} {\bibfnamefont {V.}~\bibnamefont {Ansari}},
  \bibinfo {author} {\bibfnamefont {B.}~\bibnamefont {Brecht}}, \bibinfo
  {author} {\bibfnamefont {T.}~\bibnamefont {Dirmeier}}, \bibinfo {author}
  {\bibfnamefont {C.}~\bibnamefont {Marquardt}}, \ and\ \bibinfo {author}
  {\bibfnamefont {C.}~\bibnamefont {Silberhorn}},\ }\href {\doibase
  10.1364/OE.21.013975} {\bibfield  {journal} {\bibinfo  {journal} {Optics
  Express}\ }\textbf {\bibinfo {volume} {21}},\ \bibinfo {pages} {13975}
  (\bibinfo {year} {2013})}\BibitemShut {NoStop}%
\bibitem [{\citenamefont {Midya}\ and\ \citenamefont
  {Feng}(2018)}]{PhysRevA.98.043838}%
  \BibitemOpen
  \bibfield  {author} {\bibinfo {author} {\bibfnamefont {B.}~\bibnamefont
  {Midya}}\ and\ \bibinfo {author} {\bibfnamefont {L.}~\bibnamefont {Feng}},\
  }\href {\doibase 10.1103/PhysRevA.98.043838} {\bibfield  {journal} {\bibinfo
  {journal} {Phys. Rev. A}\ }\textbf {\bibinfo {volume} {98}},\ \bibinfo
  {pages} {043838} (\bibinfo {year} {2018})}\BibitemShut {NoStop}%
\bibitem [{\citenamefont {Liu}\ \emph {et~al.}(2021)\citenamefont {Liu},
  \citenamefont {Xian}, \citenamefont {Mu}, \citenamefont {Zhao}, \citenamefont
  {Liu}, \citenamefont {Rubio},\ and\ \citenamefont
  {Wang}}]{PhysRevLett.126.066401}%
  \BibitemOpen
  \bibfield  {author} {\bibinfo {author} {\bibfnamefont {B.}~\bibnamefont
  {Liu}}, \bibinfo {author} {\bibfnamefont {L.}~\bibnamefont {Xian}}, \bibinfo
  {author} {\bibfnamefont {H.}~\bibnamefont {Mu}}, \bibinfo {author}
  {\bibfnamefont {G.}~\bibnamefont {Zhao}}, \bibinfo {author} {\bibfnamefont
  {Z.}~\bibnamefont {Liu}}, \bibinfo {author} {\bibfnamefont {A.}~\bibnamefont
  {Rubio}}, \ and\ \bibinfo {author} {\bibfnamefont {Z.~F.}\ \bibnamefont
  {Wang}},\ }\href {\doibase 10.1103/PhysRevLett.126.066401} {\bibfield
  {journal} {\bibinfo  {journal} {Phys. Rev. Lett.}\ }\textbf {\bibinfo
  {volume} {126}},\ \bibinfo {pages} {066401} (\bibinfo {year}
  {2021})}\BibitemShut {NoStop}%
\bibitem [{\citenamefont {Skirlo}\ \emph {et~al.}(2014)\citenamefont {Skirlo},
  \citenamefont {Lu},\ and\ \citenamefont {Solja\ifmmode \check{c}\else
  \v{c}\fi{}i\ifmmode~\acute{c}\else \'{c}\fi{}}}]{Skirlo2014}%
  \BibitemOpen
  \bibfield  {author} {\bibinfo {author} {\bibfnamefont {S.~A.}\ \bibnamefont
  {Skirlo}}, \bibinfo {author} {\bibfnamefont {L.}~\bibnamefont {Lu}}, \ and\
  \bibinfo {author} {\bibfnamefont {M.}~\bibnamefont {Solja\ifmmode
  \check{c}\else \v{c}\fi{}i\ifmmode~\acute{c}\else \'{c}\fi{}}},\ }\href
  {\doibase 10.1103/PhysRevLett.113.113904} {\bibfield  {journal} {\bibinfo
  {journal} {Physical Review Letters}\ }\textbf {\bibinfo {volume} {113}},\
  \bibinfo {pages} {113904} (\bibinfo {year} {2014})}\BibitemShut {NoStop}%
\bibitem [{\citenamefont {Qi}\ \emph {et~al.}(2021)\citenamefont {Qi},
  \citenamefont {Xing}, \citenamefont {Zhao}, \citenamefont {Liu},
  \citenamefont {Zhang}, \citenamefont {Hu},\ and\ \citenamefont
  {Wang}}]{PhysRevB.103.085129}%
  \BibitemOpen
  \bibfield  {author} {\bibinfo {author} {\bibfnamefont {L.}~\bibnamefont
  {Qi}}, \bibinfo {author} {\bibfnamefont {Y.}~\bibnamefont {Xing}}, \bibinfo
  {author} {\bibfnamefont {X.-D.}\ \bibnamefont {Zhao}}, \bibinfo {author}
  {\bibfnamefont {S.}~\bibnamefont {Liu}}, \bibinfo {author} {\bibfnamefont
  {S.}~\bibnamefont {Zhang}}, \bibinfo {author} {\bibfnamefont
  {S.}~\bibnamefont {Hu}}, \ and\ \bibinfo {author} {\bibfnamefont {H.-F.}\
  \bibnamefont {Wang}},\ }\href {\doibase 10.1103/PhysRevB.103.085129}
  {\bibfield  {journal} {\bibinfo  {journal} {Phys. Rev. B}\ }\textbf {\bibinfo
  {volume} {103}},\ \bibinfo {pages} {085129} (\bibinfo {year}
  {2021})}\BibitemShut {NoStop}%
\end{thebibliography}%

\end{document}